# Unwanted Advances in Higher Education: Uncovering Sexual Harassment Experiences in Academia with Text Mining

*Amir Karami[a*], Cynthia Nicole White[b], Kayla Ford[c], Suzanne Swan[b], Melek Yildiz Spinel[b]*

[a]*College of Information and Communications*
[b]*Department of Psychology*
*University of South Carolina*

[c]*Department of Psychology*
*University of Arkansas*

**Abstract**

Sexual harassment in academia is often a hidden problem because victims are usually reluctant to report their experiences. Recently, a web survey was developed to provide an opportunity to share thousands of sexual harassment experiences in academia. Using an efficient approach, this study collected and investigated more than 2,000 sexual harassment experiences to better understand these unwanted advances in higher education. This paper utilized text mining to disclose hidden topics and explore their weight across three variables: harasser gender, institution type, and victim's field of study. We mapped the topics on five themes drawn from the sexual harassment literature and found that more than 50% of the topics were assigned to the unwanted sexual attention theme. Fourteen percent of the topics were in the gender harassment theme, in which insulting, sexist, or degrading comments or behavior was directed towards women. Five percent of the topics involved sexual coercion (a benefit is offered in exchange for sexual favors), 5% involved sex discrimination, and 7% of the topics discussed retaliation against the victim for reporting the harassment, or for simply not complying with the harasser. Findings highlight the power differential between faculty and students, and the toll on students when professors abuse their power. While some topics did differ based on type of institution, there were no differences between the topics based on gender of harasser or field of study. This research can be beneficial to researchers in further investigation of this paper's dataset, and to policymakers in improving existing policies to create a safe and supportive environment in academia.

## 1. Introduction

The U.S. Equal Employment Opportunity Commission (n.d.) and U.S. Department of Education (2008) proposed the following definitions for sexual harassment, respectively:

- Sexual harassment is "unwelcome sexual advances, requests for sexual favors, and other verbal or physical harassment of a sexual nature."
- Sexual harassment "(1) is sexual in nature; (2) is unwelcome; and (3) denies or limits a student's ability to participate in or benefit from a school's education program."

Sex discrimination is a broad term that includes any unfavorable behavior in the workplace due to someone's sex, such as not hiring them, paying them less, giving them inferior work assignments, not promoting them, and so forth. Sexual harassment is a form of sex discrimination. Federal regulations in the United States, such as Title VII of the Civil Rights Act in 1964 and Title IX of the Education Amendment in 1972, established policies to prevent gender discrimination and sexism in education ("Overview of Title IX", 2015); however, sexist and discriminatory behaviors are still reported in various educational settings, especially in higher education (e.g. Biggs, Hawley, & Biernat, 2017; Settles et al., 2013). Surveys have shown that 59% of US women have experienced sexual harassment, and women with at least some college education are far more likely than those with less education to say they have experienced harassment (Graf, 2018). Regardless of status at the university (faculty, staff, or student), many women have reported that their school had a climate of sexism (Vaccaro, 2010).

In a report by the US National Academies of Science, Engineering, and Medicine survey of academic environments, 50% of female faculty/staff and 20-50% of female students reported sexual harassment experiences (National Academies of Sciences, Engineering, and Medicine, 2018). This report found that sexual harassment is an enduring problem with negative professional outcomes, such as declines in job satisfaction and productivity for faculty and staff, dropping classes and receiving lower grades for students, and psychological effects such as depression, stress, and anxiety (National Academies of Sciences, Engineering, and Medicine, 2018). Additional mental health impacts of sexual harassment include: post-traumatic stress disorder (PTSD) symptoms, psychological distress (e.g., anxiety and depression), low self-esteem, panic disorder (Petrak & Hedge, 2002; Jussen, Lagro-Janssen, Leenders, Logie, & Mijdam, 2019), and physical impacts such as nausea, headaches, fatigue, insomnia, respiratory infections, weight loss, and gastrointestinal problems (Thakur & Paul, 2017). Furthermore, sexual harassment negatively impacts the victim's work experience by leading to lower job satisfaction, lower organizational commitment, withdrawing from work, and lower worker productivity (Willness, Steel, & Lee, 2007; Lengnick-Hall, 1995). In the context of academia, women in the academic sciences, engineering, and medicine who experienced sexual harassment reported giving up tenure opportunities, leaving major research projects, or passing up leadership opportunities to avoid the perpetrator (National Academy of Sciences 2018).

Sexual harassment also has adverse financial impacts on higher education institutions. For example, the higher education insurance group paid out $36 million between 2006 and 2010 for 262 sexual harassment cases (Keehan, 2011). Sexual harassment cases cost $591,050, $1.3 million, and $10.5 million for the University of Wisconsin-Madison in 2008 (Herzog, 2018), University of Connecticut in 2014 (Nelson, 2015), and public university systems with schools in the nation's five major athletic conferences in 2016 and 2017 (Korn, 2018a), respectively. Furthermore, in 2010, the US Equal Opportunity Commission resolved over 12,000 sexual harassment cases, at a cost of over $48 million in monetary benefits to plaintiffs (McDonald, 2012). Ultimately, the negative psychological, physical, and work-related consequences of sexual harassment stifle individuals' advancement and security, as well as cause organizational financial losses (Shaw, Hegewisch, Phil, & Hess, 2018).

Although sexual harassment is widespread, it is often an invisible problem, because it so frequently occurs in situations where the only witnesses present are the victim and perpetrator. It has been difficult to combat because victims are often reluctant to report, fearing they will jeopardize their jobs or suffer other negative consequences. However, web platforms have provided an opportunity for victims to share their experiences on the Internet and social media. Online social movements, which often incorporate victims telling their stories, have radically shifted the national discourse about sexual harassment. For example, #metoo (https://twitter.com/hashtag/MeToo?src=hash), which has been used more than 19 million times between Oct 2017 and September 2018 (Geiger, 2018) on Twitter, has created a strong movement for sharing personal sexual harassment experiences (Gluckman, Read, Mangan, and Quilantan, 2017).

Many of the studies examining sexual harassment in academia are limited in sample size, and few studies examine how sexual harassment may be affected based on whether the victim is an undergraduate or graduate student, faculty, or staff; the gender of the harasser; field of study; or type of institution (e.g., Research 1 University, Liberal Arts College, etc.). In the present study, we were able to address these limitations by utilizing a sexual harassment crowdsource survey on theprofessorisin.com, an academic mentoring website (Kelsky, 2017). The survey was posted on the website and provided a place for victims in academia to share their sexual harassment experiences. More than 2,000 website users anonymously entered their stories, along with other information such as their discipline, type of academic institution, and gender of the harasser. The survey was widely publicized by different news agencies such as The Wall Street Journal (Korn, 2018b), The Guardian (Batty & Davis, 2018), and The Chronicle of Higher Education (Kelsky, 2018).

The current study adds a new perspective to the literature by analyzing posts from Kelsky's survey to get a fuller and more nuanced understanding of sexual harassment experiences in academic environments. The current study adds to the literature by tapping into numerous fields of study and types of institutions, and including experiences across academia for undergraduate and graduate students, junior and senior faculty, and staff.

We developed a mixed method approach, using both computational and qualitative methods (cf. Karami, Swan, White, & Ford, 2019). The computational approach uses text mining methods that allow researchers to analyze massive datasets by recognizing patterns and uncovering hidden knowledge in a corpus (Conte et al., 2012; Hotho et al., 2005; Karami, 2017; Karami, 2019). After completing text mining, we adopted a qualitative research method to interpret the results of text mining. Thus, the goals of the study are 1) to detect and analyze the discussion topics in users' stories and (2) to understand whether there is a difference between the topics based on the harasser gender, institute, and victims' field of study.

## 2. Related work

In Fitzgerald and Cortina's (2018) comprehensive review of research on sexual harassment, three broad categories of sexually harassing conduct have been delineated: gender harassment, unwanted sexual attention, and sexual coercion. As Fitzgerald and Cortina noted, gender harassment expresses "insulting, degrading, or contemptuous attitudes about women" (Fitzgerald & Cortina, 2018). Gender harassment is not aimed at sexual cooperation; rather, the goal is to reinforce the inferior status of the gender being targeted (Leskinen & Cortina, 2014). Subcategories of gender harassment include sexist hostility, sexual hostility, and work/family policing (2018). Sexist hostility consists of jokes, insults, and sexist comments that express negative views of women in a non-sexual way. In contrast, sexual hostility involves sexualized insults, such as referring to women by degrading names of female body parts, displaying pornographic images, or making crude comments about female sexuality (Stark, Chernyshenko, Lancaster, Drasgow, & Fitzgerald, 2002). Work/family policing includes comments that women with children are undependable students, and are not serious about their careers (Crosby, Williams, & Biernat, 2004). Gender harassment is the most widespread form of sexual harassment (Fitzgerald & Cortina, 2018; Leskinen, Cortina, & Kabat, 2011).

The second major category, unwanted sexual attention, refers to sexual advances that are uninvited and unwelcome. Such behaviors range from asking for dates, comments about someone's body or attractiveness, attempts to establish a dating or sexual relationship with someone, unwanted touching, to sexual assault and rape (Fitzgerald & Cortina, 2018). The final major category, sexual coercion, entails sexual advances in which the employee or student is offered a benefit for acquiescing (for example, a good grade or recommendation), or is threatened with a negative consequence if they do not (Fitzgerald & Cortina, 2018). These latter two categories differ in that unwanted sexual attention is not explicitly linked to a benefit, whereas in sexual coercion a link to a benefit for complying with the unwanted sexual request is implied or stated.

Researchers have investigated sexism and sexual harassment in academia. Eagly and Karau (2002) propose role congruity theory as a way to examine the prejudice women experience when entering a traditionally male-dominated role. According to Eagly and Karau (2002), discrimination against women is most likely in situations where traditional gender roles are challenged, such as when women enter Science, Technology, Engineering, or Mathematics (STEM) fields (e.g. Katila & Meriläinen, 1999). As STEM is traditionally male dominated, this theory suggests women experience prejudice because the qualities needed to be in this field are typically attributed to males rather than females. Thus, women who are pursuing these fields challenge gender stereotypes and are subjected to sexism.

The power imbalance between students and professors also plays a large role in the occurrence of harassment on college campuses. Undergraduate students are particularly vulnerable, especially earlier in their studies when they have just transitioned into the unfamiliar university setting. Despite greater experience with academia, graduate students are also not immune to being victims of harassment. A study of over 500 graduate students at a large Pacific-Northwestern public university found that 38% of female and 23% of male students reported being sexually harassed by faculty or staff (Rosenthal, Smidt, and Freyd, 2016).

Even after successful completion of graduate school and attainment of an academic job, women still encounter sexism and discrimination (Monroe and Chiu 2010). In studies of science and engineering faculty, women who perceived a sexist climate in their departments experienced more sexual harassment and were less satisfied with their jobs (Settles et al. 2006, 2012). Furthermore, while academic conferences are crucial for researchers to engage others in their work and showcase their research, socialization in this setting can promote sexist and inappropriate behaviors towards women (Biggs, Hawley, & Biernat, 2017), causing them to be less engaged in conference activities (Hinsley et al., 2017). A sexist climate at conferences also has been found to be positively related to women's intentions to exit academia altogether (Biggs, Hawley, & Biernat, 2017). Conferences can often set the tone for the field, so if a climate of sexism and harassment is established there, that is what attendees may consider the norm for the discipline; these sexist norms may then be reenacted in their own institutions (Biggs, Hawley, & Biernat, 2017).

Another hurdle experienced by those in academia is known as contrapower sexual harassment, which is when individuals with less power (i.e. students) in an organizational setting harass those with more power (i.e. professors; Benson, 1984). DeSouza (2011) conducted a survey with university faculty and found that 22% experienced sexual harassment from a student. These studies expose the reality that despite the power of a faculty position, professors are not immune to harassment from students.

Not only do faculty, staff, and students experience various forms of sexism and harassment on campus, but they also often do not receive support when they try to address the problem. Goltz conducted a qualitative study with female college students and faculty who reported experiencing sex discrimination at their American university (2005). The women were in a wide range of fields, and the discrimination ranged from unequal pay to unequal promotion/hiring to sexual harassment. Participants first tried to informally address the discrimination by speaking to administrators, colleagues, or professional organizations. However, informal appeals rarely lead to helpful conclusions, and even when a formal appeal followed, that was often met with denial of responsibility, inaction, or retaliation (Glotz, 2005). While the sample size was small, this study suggests that victims' attempts to end harassment and sexism are rarely successful and may make the situation worse. A study of sexual harassment among students in medical school revealed similar concerns about retaliation (Wear & Altman, 2005). As one participant in this study remarked, "Don't bring it [sexual harassment] up because it's going to hurt you in the end, it's better to stay quiet, not say anything, let it happen, take your grade at the end, be thankful that you passed" (p. 5).

In sum, the literature suggests that harassment is still present and pervasive in academia. However, the existing literature on sexual harassment and sexism in academia is limited in scope, and many studies have relatively small sample sizes. There is a need to better understand the sexual harassment experiences and patterns in academia (Seto, 2019).

## 3. Methodology

In this section we describe the construction of the corpus and data analysis methods utilized in this research.

### 3.1 Data

We collected 2,379 sexual harassment experiences in academia from the sexual harassment crowdsource survey on theprofessorisin.com website. This survey asked the users for information about their personal sexual harassment stories in academia, along with some other information such as the type of institution where the sexual harassment took place. The dataset and its meta-data in this research are available at https://github.com/amir-karami/Academia_Sexual_Harassment.

We chose to analyze Kelsky's survey data over social media data such as #metoo tweets for several reasons. First, in the survey, users did not have any restrictions on the length of their stories such as the number of characters limit on Twitter; therefore, users had enough space to provide the details of their experiences. Second, the entries had other information such as the victims' field of study. This information helped us to add more dimensions to our analysis. Third, as the focus of this paper is on sexual harassment in academia specifically, #metoo tweets would require a pre-processing step to cluster the #metoo tweets in different sexual harassment categories, such as workspace and academia. The accuracy of this process has some errors and potential data loss. However, the survey used for the present study doesn't need clustering because it has been specifically developed for sexual harassment in academia.

## 3.2 Text mining

In this research, we used two text mining techniques: frequency analysis to provide an impression of the corpus and topic modeling to discover hidden semantic structure of the academic sexual harassment stories. In the frequency analysis, we utilized word cloud visualization, with larger font size indicating higher frequency (Karami, Ghasemi, Sen, Moraes, & Shah, 2019). Although frequency analysis gives a basic perspective, this analysis does not extract hidden semantic structure of a corpus. Therefore, we need advanced text mining methods to discover new semantic layers.

Different text mining methods have been proposed, with topic modeling as a popular method to discover topics in a corpus (Karami, Gangopadhyay, Zhou, & Kharrazi, 2015; Karami, Gangopadhyay, Zhou, & Kharrazi, 2018). Among different topic models, latent Dirichlet allocation (LDA) (Beli et al., 2003) is a popular model that has superior performance over other similar methods such as co-occurrence analysis and latent semantic analysis (Lee, Song, & Kim, 2010; Sugimoto, Li, Russell, Finlay, & Ding, 2011).

LDA is a generative probabilistic model that assigns words that occurred together in a corpus to a category called topics (Beli et al., 2003). LDA assumes that the words in a topic are semantically related and represent a theme (Karami, 2015). For example, LDA assigned the words "conference," "room," "hotel," "dinner," "senior," "scholar," "research," "academic," "reception," and "night" to a topic. Using qualitative methods described below, we interpreted this topic as "Harassment & assault by male faculty at conferences. Men thinking conferences are a free pass" (Topic 31, Table 1). This topic model has been used for different applications such as health (Shaw & Karami, 2017; Zhu, Kim, Banerjee, Deferio, Alexopoulos, & Pathak, 2018; Karami, Webb, & Kitzie, 2018; Webb, Karami , & Kitzie 2018; Karami, Dahl, Turner-McGrievy, Kharrazi, & Shaw, 2018; Karami & Shaw, 2019), e-petition (Hagen, 2018), politics (Park, Chung, & Park, 2019; Karami, Bennett, & He, 2018; Karami & Elkouri, 2019), opinion mining (Ma, Zhang, Liu, Li, & Yuan, 2016), disaster management (Karami Shah, Vaezi, & Bansal, 2019) business (Amado, Cortez, Rita, & Moro, 2018; Karami & Pendergraft, 2018), social media analysis (Karami & Collins, 2018; Collins & Karami, 2018), automatic summarization of changes in dynamic text collections (Kar, Nunes & Ribeiro, 2015), spam detection (Karami & Zhou, 2014), and systematic literature review (Wang, Ding, Zhao, Huang, Perkins, Zou, & Chen, 2016; Altena, Moerland, Zwinderman, & Olabarriaga, 2016; Karami et al., 2019; Shin et al., 2019). We utilized LDA in this research to achieve a deeper semantic layer in the academic sexual harassment corpus. To discover the meaning of the topics, we employed a qualitative approach for coding the topics.

The outputs of LDA for n documents (experiences), m words, and t topics, are two matrices. The first one is the probability of each word in each topic or $P(W_i|T_k)$ and the second one is the probability of each topic in each document or $P(T_k|D_j)$:

$$Words \begin{bmatrix} P(W_1|T_1) & \cdots & P(W_1|T_t) \\ \vdots & \ddots & \vdots \\ P(W_m|T_1) & \cdots & P(W_m|T_t) \end{bmatrix} \quad \& \quad Topics \begin{bmatrix} P(T_1|D_1) & \cdots & P(T_t|D_n) \\ \vdots & \ddots & \vdots \\ P(T_t|D_1) & \cdots & P(T_t|D_n) \end{bmatrix}$$

$$P(W_i|T_k) \qquad\qquad\qquad\qquad P(T_k|D_j)$$

The top words in each topic based on the descending order of $P(W_i|T_k)$ represent each of the topics. Interpreting the top words of a topic is part of the information used to interpret the overall theme of topic. On the other side, the most related documents of the topic based on the descending order of $P(T_k|D_j)$ can help to better understand the topic. We also used $P(T_k|D_j)$ to find the weight or significance of each topic, $ST(T_k)$. To have an effective comparison, each of the STs was normalized by the sum of the weight scores of all topics $N\_ST(T_k)$.

$$N\_ST(T_k) = \frac{\sum_{j=1}^{n} P(T_k|D_j)}{\sum_{k=1}^{t} \sum_{j=1}^{n} P(T_k|D_j)}$$

If $N\_ST(T_x) > N\_ST(T_y)$, it means that topic x is discussed more than topic y. $N\_ST(T_k)$ can also be used to measure the weight of a subset of topics. In this research, we explore the relationship between weight of the topics by three variables: harasser gender, institution type, and victim's field of study.

## 3.3 Qualitative topic analysis

To disclose the meaning of topics and their categories, we implemented a qualitative approach in four phases: (1) discovering the theme for each topic, (2) detecting the relevant and meaningful topics, (3) determining the overarching categories, and (4) assessing reliability of coding. We explain each of these phases below.

**Phase 1: Discovering the theme for each topic**. To make the determination of the topics, three of the authors coded the topics individually. "Coding" in this context means that coders read the top 10 words (shown in Table 1) and top 10 stories for each of the topics and tried to identify the common theme underlying the stories. To find the top stories for each topic, we sorted $P(T_k|D_j)$ from the highest value to the lowest one. The three coders used consensus coding to agree on the theme for each topic, using Lim, Valdez, & Lilly's (2015) consensus coding method. For consensus coding, the coders first developed themes separately; then they met, compared and contrasted the themes they had each generated, and kept on discussing them until they agreed on the final themes. For example, one topic contained these words: "*comments, made, sexual, inappropriate, jokes, remarks, touching, unwanted, sexually, uncomfortable, lewd, making, and repeated.*" After each of them coded these topic words and its corresponding top 10 stories individually, they came together, discussed it, and reached consensus on coding this topic as "Sexual Remarks & Touching" (see Topic 30, Table 1).

**Phase 2: Detecting the relevant and meaningful topics**. The next step was to determine the topics that were meaningful or directly related to academia, culling topics that did not fit the aims of the study. Again, we used the consensus coding method, refined the topics, and agreed on 41 relevant and meaningful topics. The four topics that were removed were related to news coverage about sexual harassment rather than website users' personal stories, or contained multiple issues and did not have a consistent theme.

**Phase 3: Determination of overarching categories**. In the final phase, coders grouped the 41 topics into the themes and subthemes shown in Table 2. Again using consensus coding, we grouped the topics, guided by the four types of sex discrimination and sexual harassment identified in the literature: (a) sex discrimination; and the three types of sexual harassment, consisting of (b) gender harassment (with the three subtypes [sexist hostility, sexual hostility, work/family policing]); (c) unwanted sexual attention; and (d) sexual coercion.

**Phase 4: Assessing reliability of coding.** Once final coding was completed, we utilized an outside coder who was not involved in the project to check our consensus coding. In this way, we could determine if, given the same dataset, another person would reach the same conclusions as we did regarding which topics fit into the themes and subthemes. The outside coder coded 11 of the 41 topics (27% of the total number of topics), making a determination as to which of the themes and subthemes those 11 topics fit into. Then, we performed a Cohen's κ to determine if there was agreement between our coding reached via consensus (described in Phase 1) and the outside coder. There was very good agreement, κ = .80, $p$ < .0005 (Altman, 1991).

## 4. Results

Word frequency analysis shows that 86% of the words appeared less than 10 times. With median 2 and average 7.95, the frequency of the 13,395 words in our corpus is between 1 and 1,261. Figure 1 is in line with Zipf's law and illustrates the position of the top 50 words among the top 1000 high frequent words. Zipf's law states that the frequency of a word in a corpus is inversely proportional to its frequency rank (Zipf, 1949). Figure 2 shows the word cloud of the top-50 high frequency words. The words "male" and "department" are the most frequently used words, followed by "faculty," "class," and "comments." "Grad" is the next most frequently used word, indicating the substantial proportion of survey respondents who wrote about sexual harassment in graduate programs.

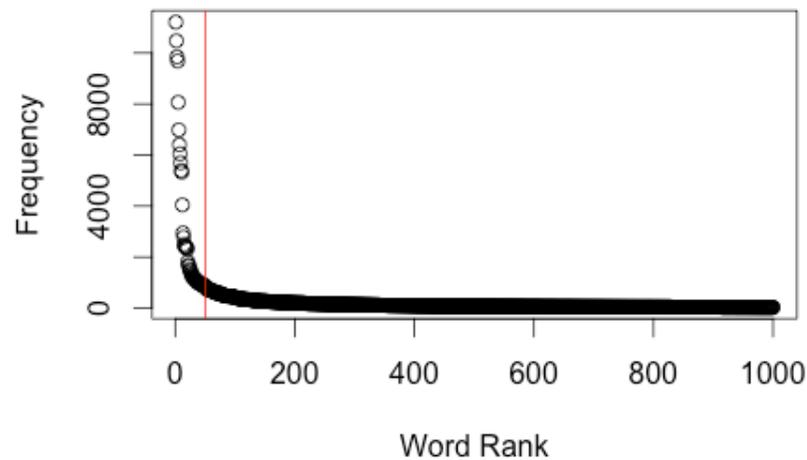

**Figure 1**: Frequency of Words. The vertical line shows cut-off point for the top-50 words in the word cloud.

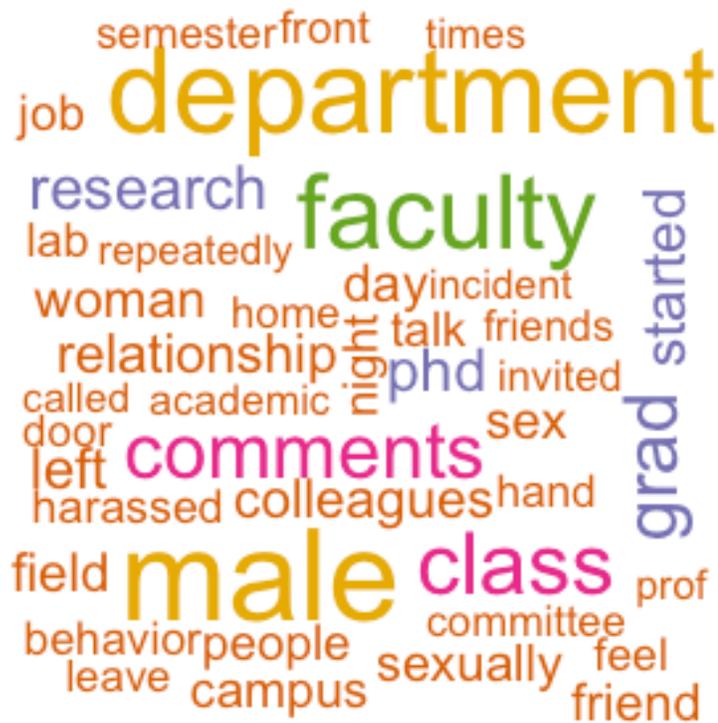

Figure 2: Word Cloud of the Top-50 Words

To detect the optimal number of topics, we applied a density-based method that assumes that the best performance of LDA is at the minimum average cosine distance of topics (Cao et al., 2009). Applying *ldatuning* R package[1] on the number of topics from 5 to 300 showed the appropriate number of topics at 45.

To discover the 45 topics in the corpus, we applied the MALLET implementation of LDA (McCallum, 2002) that was developed based on Java programming language for text mining purposes. This step comes with removing the stopwords such as "the" and "a" that do not have semantic value for our analysis. Then, we evaluated the robustness of LDA using the log-likelihood for five sets of 1000 integrations. Training the MALLET on the five sets showed that LDA reached its maximum value before 1000 iterations (Figure 3). We compared the five iterations and found that there was not a significant difference (P-value >0.05) between the five iterations with respect to mean and standard deviation. Then, we applied the MALLET with 1000 iterations and 45 topics. Using N_ST(T$_k$) shows that the weight of topics ranged from 0.0165 to 0.0381 with average 0.02 (Figure 4).

---

[1] https://cran.r-project.org/web/packages/ldatuning/vignettes/topics.html

## 4.1 Themes and subthemes

We removed T9, T19, T23, and T24 because they were not meaningful or relevant topics. Table 1 shows the 41 meaningful and relevant topics along with their description and weight ranking. There were five major themes that emerged from our data: gender harassment, sex discrimination & harassment, unwanted sexual attention, sexual coercion, and retaliation (see Table 2). The first four themes were consistent with the sexual harassment literature, as reviewed above (EEOC, n.d.; Fitzgerald & Cortina, 2018; Karami et al., 2019). The final theme, retaliation, was not a type of sexual harassment per se. Rather, this theme emerged as website users posted stories in which they experienced retaliation for reporting harassment to authorities, or simply for not complying with the harasser. Below, we describe the results from the website and included quotations from exemplar stories for each of the topics (see Table 3). The quotations in the table are presented in their original form to uphold integrity of the stories; therefore, there may be grammar and spelling errors.

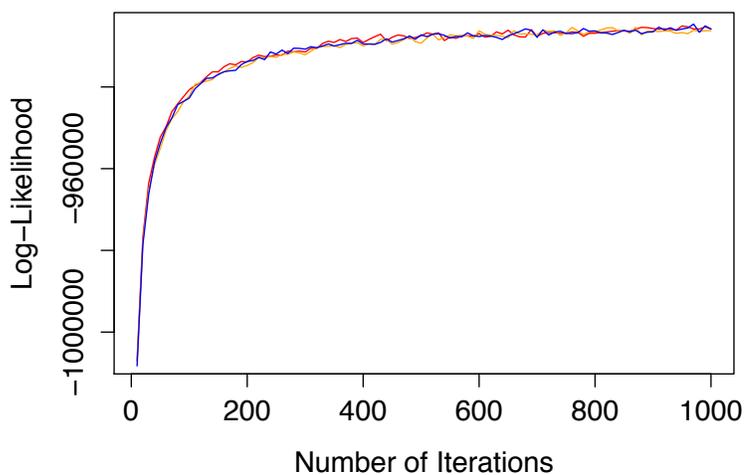

Figure 3: Convergence of the Log-Likelihood for 5 sets of 1000 iterations

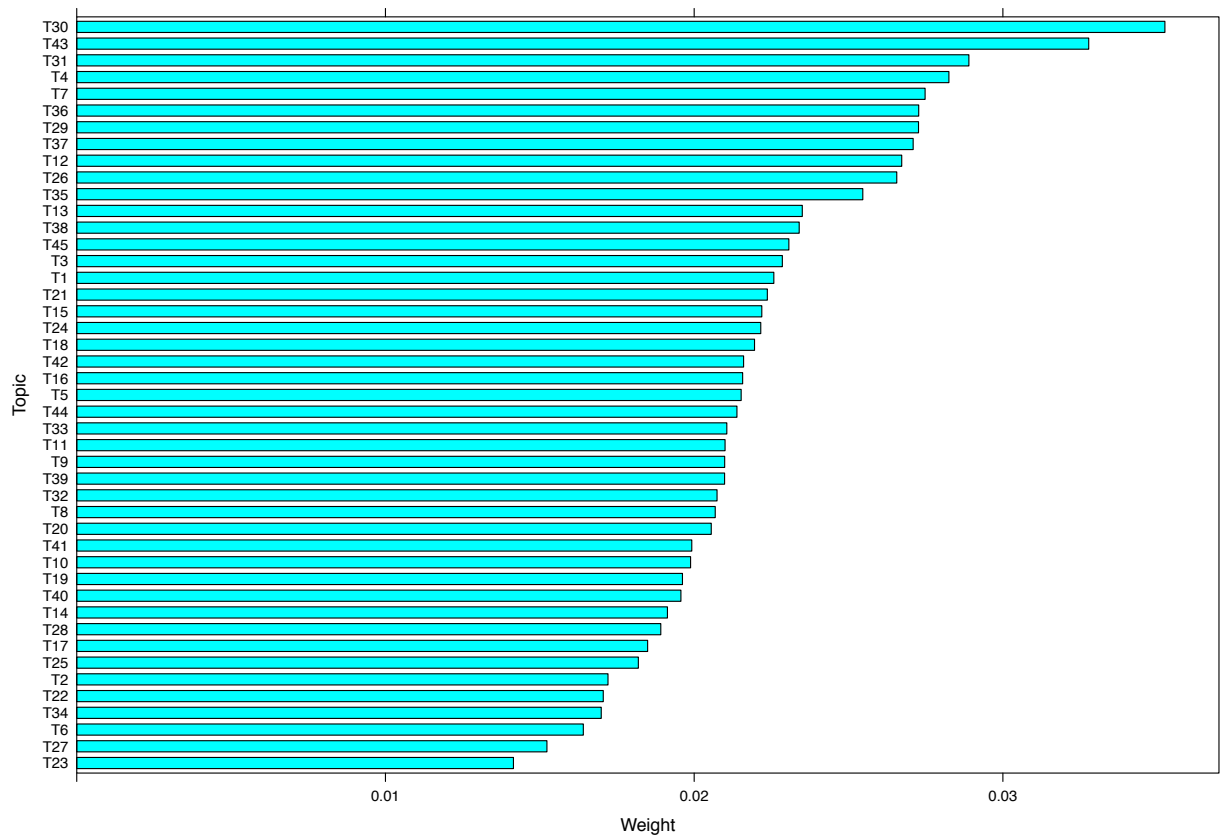

Figure 4: Ranked Weight of Topics

Table 1. Topics, description, words comprising each topic, and weight ranking of topic (higher weighted topics are discussed more in the data). The 10 top weighted topics appear in bold.

| Topic# | Description | Words/Topic | Weight Ranking |
|---|---|---|---|
| T1 | Attempts to turn professional interactions into romantic or sexual interactions | room dinner hotel meeting home thought car colleague insisted meet | 16 |
| T2 | Sexual Bullying, threatening, retaliation for reporting | university campus student police story case committee threats involved perpetrator | 40 |
| T3 | Unwanted touching and suggestive behavior and comments | back room sat walked conversation door talking couch looked hand sitting | 15 |
| **T4** | **Harassment at all levels - undergraduate, graduate, faculty** | **professor students department graduate student undergraduate things harassed friend sexually** | **4** |
| T5 | Graduate students threatened or harassed by faculty. Retaliation for not complying or speaking up. | advisor program committee phd dissertation thesis graduate made career needed | 23 |
| T6 | Inappropriate behavior or sexual bullying | people wanted called face institution men place put language idea | 43 |
| **T7** | **Professor getting student alone so he can harass her** | **felt back uncomfortable make university started left moved eventually touch** | **5** |
| T8 | Harassment & sexism involving grad students & postdocs as both victim & perpetrator | lab research student postdoc group mentor working project assistant uncomfortable | 30 |
| T10 | Professors initiating personal conversations about sex | sex life personal rumors questions professor older penis talking private | 33 |
| T11 | Harassment & sexual advances at off campus social events | professor wife invited summer home house party lunch undergraduate visit | 26 |
| **T12** | **Sexual violence & threats** | **sexually harassed assaulted raped physically stalked threatened repeatedly verbally attempted** | **9** |
| T13 | Treating work environment as sexual supermarket | student graduate department fellow program sit events social attended dating | 12 |
| T14 | Professors using power to proposition students or influence their success | professor paper final wanted thought suggested semester point exam grade | 36 |
| T15 | Using meetings to try to make mentoring relationship sexual | work research meeting project talk long advice coffee gave invited | 18 |
| T16 | Professors trying to manipulate students into sexual relationships | relationship work began friend academic mentor ended marriage wife adviser | 22 |
| T17 | Discrimination towards mothers or pregnant women | pregnant married big husband children career men young child baby | 38 |
| T18 | Student needs letter from professor but instead gets propositioned | school graduate professor student letter recommendation thesis write applying relationship | 20 |
| T20 | Sexual comments & unwanted touching | man back started thought talking felt hard continued make | 31 |
| T21 | Attempts to report harassment that resulted in retaliation | department chair dean reported office hr complaint behavior harassment filed | 17 |
| T22 | Music & art department harassment as norm | program major teacher incident high entire occurred knew voice arts | 41 |
| T25 | Inappropriate sexual behavior at outside professional activities | group event dinner proceeded immediately drink work home sleep car | 39 |
| **T26** | **Senior faculty using harassment to assert power** | **faculty member senior department members tenured junior chair person repeatedly** | **10** |
| T27 | Bystander complacency with harassment | director team boss worked stop order end approach left felt | 44 |
| T28 | Harassment & sexist bullying from someone with power | phd supervisor university work started behaviour left finished research offered | 37 |
| **T29** | **Comments on body & appearance in professional setting** | **comments made looked inappropriate wearing appearance body uncomfortable breasts sexy** | **7** |
| **T30** | **Sexual remarks & touching** | **comments made sexual inappropriate jokes remarks touching unwanted uncomfortable lewd** | **1** |
| **T31** | **Harassment & assault by male faculty at conferences. Men thinking conferences are a free pass.** | **conference room hotel dinner senior scholar research academic reception night** | **3** |
| T32 | Predatory professors, harassment in classroom | students sexual graduate multiple advances class abuse reputation regularly inappropriate | 29 |
| T33 | Repeated harassment by male professors misusing their power | work good students gave knew guy find felt writing needed | 25 |
| T34 | Expected to socialize or have sexual relationship with to be included in important scholarly activities | sexual field harassment career heard story stop early experience involved | 42 |

| T35 | Professors harassing students. Often known to department but nothing is done. | grad student program professor undergrad school fellow found early reported | 11 |
|------|------|------|------|
| **T36** | **Professors encouraging drinking and using settings with alcohol to make advances.** | **night friends bar party left apartment wanted drink invited kissed** | **6** |
| **T37** | **Sexualizing the classroom environment** | **class professor students ta teaching front grade end taking classroom** | **8** |
| T38 | Department Chair or Dean setting the tone for systemic sexism & harassment | department job position chair interview head dean hire give offer man | 13 |
| T39 | Professors coercing or grooming students into sexual relationships | relationship sex sexual situation consensual romantic affair friends dating emotional | 28 |
| T40 | Continuum of Inappropriate sexual behavior to rape & stalking | night happened called late campus left times previous semester sleep | 35 |
| T41 | Male professors using international or remote field sites to harass students | professor students trip sleep comments site sexist fieldwork camp summer | 32 |
| T42 | Offensive comments & aggressive sexist bullying | colleague tenure senior chair prof track assistant committee students inappropriate | 21 |
| **T43** | **Unwanted touching, groping by professors** | **hand put professor back grabbed touching kiss thigh shoulder arm** | **2** |
| T44 | Inappropriate office behavior such as using porn that others see | office door closed hours working heard day computer porn close | 24 |
| T45 | Unwanted persistent sexual messages by males | email contact messages began phone text telling leave call message | 14 |

Table 2. Themes, Subthemes, Topics comprising each Subtheme, and Description of Subtheme

| Theme | Subtheme | Topic(s) | Description |
|---|---|---|---|
| **Gender Harassment** | Gender Harassment | T6, T26, T27, T30 T33, T44 | Inappropriate office behavior that combines a sexualized environment with sexist statements that express contempt for women. Using harassment to assert power; thus, creating a hostile environment. |
| | Sexist & Sexual Hostility | T37, T42 | Offensive comments (i.e. sexist jokes or referring to women by degrading names of female body parts) and aggressive sexist bullying |
| | Work/Family Policing | T17 | Discrimination towards mothers or pregnant women. |
| **Sex Discrimination & Harassment** | | T8, T38 | Department Chairs or Deans set the tone for systemic sexism & harassment. Graduate students and postdocs involved as both victims and perpetrators of harassment. |
| **Unwanted Sexual Attention** | Sexual Supermarket | T4, T7, T10, T13, T15, T18, T20, T29, T32 | Professors see the academy as their sexual supermarket, picking students to pursue like they're choosing apples at the market. This includes sexual comments, unwanted touching, or attempts to date students. The professors model sexual harassment and set the tone for what is normative in the department. This results in an environment in which people with more power feel they have a free pass to harass those with less power: Professors hit on students, senior faculty hit on junior faculty, graduate students and post-docs hit on younger graduate students and undergraduates, etc. |
| | Unwanted touching | T3, T43 | Unwanted touching, groping, grabbing |
| | Professor- Student Affairs | T16, T34, T35, T39 | Professors have affairs with students that are characterized by a marked power difference between the professor and student. This includes abusive professors who "groom" students into affairs, then abuse them. The affair is "consensual" in the sense that the student makes a "choice" to have an affair with the professor. |
| | Off-Campus Sexual Advances | T1, T11, T25, T31, T36, T41 | Harassment & sexual advances at off campus events such as conferences. Professors treat conferences as a free pass to behave in ways typically deemed inappropriate for the office. Professors encourage drinking and use settings with alcohol to make advances. |
| | Aggressive violence | T12, T40, T45 | Continuum of inappropriate sexual behavior: threats, sexual violence, stalking, rape |
| **Sexual Coercion** | | T14, T28 | Sexual advances in which a person is offered some kind of benefit for accepting, or is threatened with a negative academic-related consequence if they do not give in. |
| **Retaliation** | | T2, T5, T21 | Retaliation for not complying with the harasser, or for speaking up or reporting the harassment. Retaliation includes sexual bullying and threats. |

### 4.1.1 Gender harassment theme

Three subthemes and nine topics were included in the theme of gender harassment. The three subthemes included gender harassment, sexist hostility and sexual hostility, and work/family policing. The first subtheme, gender harassment, included inappropriate office behavior that combines a sexualized and hostile environment with sexist statements that express contempt for women. Notably, Topic 30 from this subtheme, sexual remarks and touching, was the number one weighted topic (Table 3, S1).

Topic 26, senior faculty using harassment to assert power, was the 10th highest weighted topic and was also in this subtheme. There were several stories about professors using their power to harass those below them, as well as examples of male faculty assuming women are less capable than men. (Table 3, S2).

In the sexist hostility and sexual hostility subtheme, people experienced offensive comments (i.e. sexist jokes or referring to women by degrading names of female body parts) and aggressive sexist bullying. This aggression sometimes came from coworkers or fellow graduate students (Table 3, S3). The 8th highest weighted topic, topic 37, sexualizing the classroom environment, was also in this subtheme. This included stories about professors utilizing the classroom to sexualize women and telling women they should behave in a certain manner (Table 3, S4).

The final subtheme of gender harassment was work/family policing. Multiple respondents wrote stories about being refused promotions or raises due to their status as mothers. One woman, a staff employee, described her boss, a full professor, groping her from behind while she was pregnant and fearing that no one would believe her if she reported it because she was pregnant. Sometimes women's intelligence or commitment to academia was questioned due to motherhood (Table 3, S5 and S6).

### 4.1.2 Sex discrimination & harassment theme

This theme combines both sex discrimination and harassment[2]. In some departments, chairs or deans set the tone for systemic sexism and harassment (Table 3, S7). Because those in leadership are modeling harassment and discrimination, those below them in the hierarchy not only must tolerate this behavior, but also are encouraged to display similar behavior and come to the harasser's defense. For example, a user wrote about their male department chair's pervasive sexual harassment of female staff. Eventually, enough people complained that he was placed on administrative leave; however, other male faculty defended him. Graduate and postdoctoral students were also involved in harassment as both victims and perpetrators (Table 3, S8).

### 4.1.3 Unwanted sexual attention theme

As shown in Table 2, the unwanted sexual attention theme had the greatest number of subthemes and topics, illustrating how common this form of harassment is in academia.

---

[2] It is not surprising that some of the topics overlapped, such as sex discrimination and sexual harassment, because of the common co-occurrence of different types of discrimination and harassment (Leskinen et al., 2011).

Table 3. Example stories of the subthemes that emerged from the sexual harassment survey

| Theme | Subtheme | Story ID | Story |
|---|---|---|---|
| **Gender Harassment** | Gender Harassment | S1 | *"I was told my department "only hired the pretty [women]" It was these kinds of comments over years and years. It was using the word "pussy" in a meeting as a synonym for weakness."* |
| | | S2 | *"While in my surgical training, an attending physician broke a valuable instrument. I was called into the department chief who accused me of breaking it. I said I had not. He replied, 'Well, you were the only woman there.'"* |
| | Sexist & Sexual Hostility | S3 | *"I TA'd for a male professor along with two other members of my cohort, one man and one woman...the male TA repeatedly made "jokes" in front of our professor and students about how I and the other female TA were unqualified to teach, knew nothing about the subject matter of the class, would fail our quals, were overly emotional or stupid, etc."* |
| | | S4 | *"You should change your personality', as a woman, people will like you better."* |
| | Work/Family Policing | S5 | *"I overheard two male peers discussing how women who have children could never really ever be dedicated to science".* |
| | | S6 | *"One young mother said she "did not [get] a position for having a baby".* |
| **Sex Discrimination & Harassment** | | S7 | *"In an interview for an Assistant Dean position, I [was] wearing a suit jacket and skirt, the hiring Dean smirked, told me he was a "leg man" and patted me on my bare knee".* |
| | | S8 | *"I was hired into a lab as a postdoc with the agreement that I would learn electrophysiology, only to be directed into molecular portion of the lab over and over again. I learned that according to my PI women do not have the 'temperament' to do electrophysiology."* |
| **Unwanted Sexual Attention** | Sexual Supermarket | S9 | *"During a graduate conference, a full professor...smacked the buttocks of an assistant professor and groped a female graduate student".* |
| | | S10 | *"A full professor (female) made repeated references to my body and asked repeated questions about my sex life. She made numerous comments about her own sex life all when we were alone in her lab."* |
| | | S11 | *"Male tenured professor serially groped female students/serially said inappropriate and intimidating things to female students in the department over the course of many years and nothing was ever done about it."* |
| | Unwanted Touching | S12 | *"I was at a department social gathering and a senior member came up very close, slid their hand around my waist and then up and grabbed my breast".* |
| | | S13 | *"Academic advisor put his hand on my knee and tried to run it up my skirt"* |
| | | S14 | *"A fellow undergrad student invited me to watch a movie with him. We sat on his couch (upright, not touching)...After about 10-15 minutes, he grabbed my wrist and forced my hand onto his penis. I tried to leave the room, but he beat me to the door and held it closed."* |
| | Professor-Student Affairs | S15 | *"Relationships between faculty and grad students were normal, and so harassing behavior could be presented as normal too"* |
| | | S16 | *"I met a well-known professor...I was very keen to work with him and learn from him. So I initially approached him. We rode the metro home together from the library and he was very solicitous and friendly, so eventually I confided in him that I was having suicidal thoughts...he presented himself as a concerned person, a kind of rabbi or therapist...I now realize that he was grooming me. We were "friends" for about nine months before he made a pass at me...I trusted him. I did not think he could hurt me. What started out as sitting on the couch and talking eventually led to some very intimate sex in his bed...He told me he had had many affairs with graduate students, because he was unhappy in his marriage. He led me to believe that I was different than the ones who came before and that we had a future. Then he went back to his wife and cut things off. I became even more unhinged and depressed and suicidal."* |
| | Off-Campus Sexual Advances | S17 | *"When I was a first-year undergraduate...There was a party for the choir and chapel staff; alcohol was provided by the staff, and we were all encouraged to drink a lot. One of the assistant chaplains... aggressively and single-mindedly tried to chat me up... no one stepped in, even though they could see he was touching me extremely inappropriately and I was incapable of consent."* |
| | | S18 | *"My advisor tried to have sex with me when on a trip to South Africa. I turned him down. The next day he took a proposal away from me that I had been working on for about 1 year."* |
| | | S19 | *"A prominent man in my field grabbed my thigh while sitting next to me at a conference dinner. He later cornered me at a reception, told me his wife didn't care what happened at conferences, and invited me to back to his hotel room."* |
| | Aggressive violence | S20 | *"I was drugged and raped by a professor in my department at the beginning of this semester."* |
| | | S21 | *"Group of male fraternity students sent an email to me about needing the hottest teacher on campus to attend a party. There were many references to my appearance."* |
| | | S22 | *"Stalked by a lecturer, who on one notable occasion told me he would hurt me should I continue to refuse him; left a long and violent message on my answer machine; camped outside my house for 48 hours."* |
| **Sexual Coercion** | | S23 | *[As an undergraduate student my professor] "offered me an A if I would spend the weekend with him. I had gone to his office hours in order to get ideas for the final paper."* |
| | | S24 | *"Final semester of my masters degree program…My professor came to the final rehearsal of my grade recital…He told me that my performance was not sufficient and that I would fail the recital, thereby not earning my degree. I had a 3.9 GPA…had [been] given performance awards by this professor…When I sat down on his office sofa, in shock…he sat next to me, placed his hand on my inner thigh, and said 'but there is one way you could still pass.' I stood up, left, and performed the recital. He left a message on my phone later confirming that I would not receive the degree…I was in therapy for four years afterward, medicated, and hospitalized for suicidal ideation."* |
| **Retaliation** | | S25 | *"I reported a threat to a campus doctor who put me in touch with the campus police. The police ignored me but contacted the department chair who had made her negative feelings for me known. She then forced me to sit on a graduate committee with my harasser. She asked him for any incriminating evidence in private gmail communications… I complained again and was demoted."* |
| | | S26 | *"I witnessed a faculty member say he was going to 'destroy' a student who had filed a harassment claim against him. I reported the retaliation. He then came after me with others piling on."* |

It is not surprising, then, that 7 of the 10 highest-weighted topics are in this theme. The subthemes include: sexual supermarket, unwanted touching, professor-student affairs, off-campus sexual advances, and aggressive violence. The first subtheme, sexual supermarket, is characterized by professors seeing the academy as a place where they can pick students to pursue (Fitzgerald & Weitzman, 1990). This attitude that students are a commodity that professors have a right to pick and choose leads to behavior that can include sexual comments, unwanted touching, or attempts to date or have sex with students. The harassment by professors results in an environment in which people with more power feel they have a free pass to harass those with less power: senior faculty hit on junior faculty, professors hit on students, graduate students and post-docs hit on younger graduate students and undergraduates, etc. The sexual supermarket subtheme included 3 of the top 10 highest-weighted topics: Topic 4, harassment at all levels – undergraduate, graduate, and faculty, weighted 4th; Topic 7, professor getting student alone so he can harass her, weighted 5th; and Topic 29, comments on body and appearance in professional setting, weighted 7th (Table 3, S9-S11).

The second subtheme, unwanted touching, was focused on groping and grabbing. Topic 43, unwanted touching and groping by professors, is in this subtheme and was the second highest-weighted theme. There were many instances of unwanted touching occurring at social gatherings. This type of victimization not only occurred by professors, but peer on peer victimization was also mentioned (Table 3, S12-S14).

Professor-student affairs is the third subtheme within unwanted sexual attention. In this subtheme, professors with considerable power over students seek affairs and sex with them. Students are often groomed into these "consensual" affairs, and then abused by the professor. In some cases, affairs were so pervasive that they became the norm (Table 3, S15). Sometimes professors lured students into a relationship by suggesting that they would leave their spouse, with devastating consequences for the students (Table 3, S16).

In the third subtheme, *off-campus sexual advances*, professors treated conferences and off-campus events as a free pass to behave in ways deemed inappropriate for the office. Topic 31, harassment and assault by male faculty at conferences, the third highest-weighted topic, and topic 36, professors encouraging drinking and using setting with alcohol to make advances, the 6th highest-weighted topic, were in this subtheme. A number of stories referred to experiences at conferences, used settings with alcohol to make advances, and made sexual advances during international study abroad or fieldwork experiences (Table 3, S17-S19).

The final subtheme of unwanted sexual attention is aggressive violence. This subtheme contained Topic 12, sexual violence and threats, the 9th highest weighted topic. These stories fell along a continuum of violent sexual behavior: threats, sexual violence, stalking, and rape. One person had their scholarship threatened if they disclosed the repeated groping they endured, while other users described stalking and threats. This subtheme also contained some stories about contrapower sexual harassment (Table 3, S20-S22).

## 4.1.4 Sexual coercion theme

Some students described sexual coercion from professors. These experiences were not limited to undergraduates, but graduate and doctoral students were threatened as well. Students had to do something sexual to get a specific grade or complete a major milestone within their program. These instances of sexual coercion had a grave impact on students, and highlight the imbalance of power in academia (Table 3, S23-S24).

### 4.1.5 Retaliation

Retaliation for reporting the harassment, or for simply not complying with the harasser, is the final theme. Retaliation often included bullying and threats. In some cases, the retaliation occurred at the highest levels of the university (Table 3, S25 and S26).

### 4.2 Harasser gender, institute, and victims' field of study

After completing the topic and theme analyses, we conducted further analyses to examine if other variables in the survey's meta-data were related to the topics. The collected data contained three variables including (1) harasser gender[3], (2) the type of institution where the harassment happened, and (3) victim's field of study. Considering the first variable, we found that 97.14% of the users defined the harasser gender in four categories including male, female, non-binary, and both female and male (Figure 6). The vast majority of harassers were male (90%); a small number were female (5%). Due to the very low number of experiences in the non-binary, and the both female and male categories, we compared the female and male categories based on the average weight of the 41 topics. Our results show that there isn't a significant difference (P-value >0.05) between the topics of female and male categories.

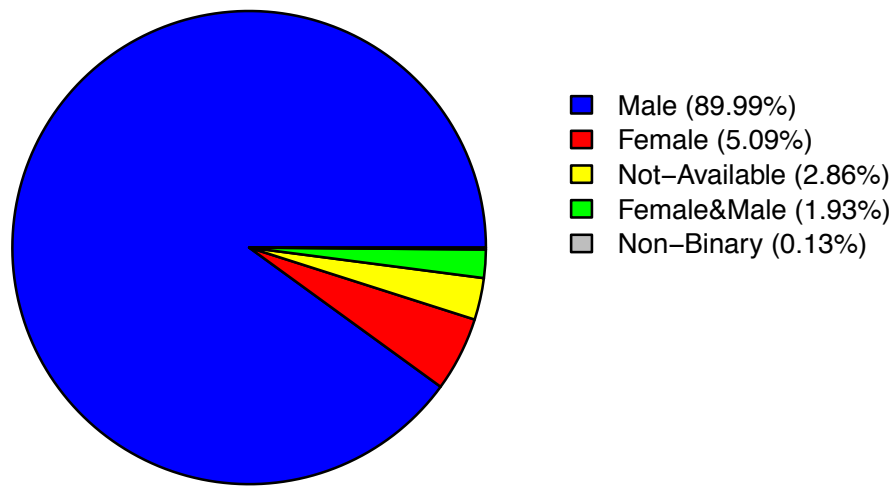

Figure 6: Distribution of Harasser Gender

The analysis of the type of institute shows that 96.93% of the users reported the types of institutes in eight categories including (1) elite institution or Ivy league (EI), (2) other research 1 (OR1), (3) research 2 (R2), (4) small liberal arts college (SLAC), (5) regional teaching college (RTC), (6) other type of school, (7) more than one institute, and (8) other research agency (Figure 8). Most of the harassment reported in the survey occurred at R1 (40%) and elite/Ivy league universities (25%), likely reflecting the population of people who used the [theprofessorisin.com](theprofessorisin.com) website. Due to the ambiguity of answers for categories 6, 7, and 8, we compared the elite institution or Ivy league, other R1, R2, small liberal arts college, and regional teaching college categories based on the average weight of the 41 topics. Our findings show that some harassment topics do differ based on type of institution. The analysis of this research illustrates a significant difference (P-value >0.05) between the elite institution/Ivy league and the R2 based on topic 32 (R2 > EI), the elite institution/Ivy league and regional teaching college based on topics 1 and 20 (RTC > EI), other R1 and R2 based on topic 32 (R2 > OR1), other R1 and regional teaching college based on topic 1 (RTC > OR1), R2 and regional teaching college based on topics 7 and 20 (RTC > R2), regional teaching college and small liberal arts college based on topics 1 and 20 (RTC > SLAC).

---

[3] The survey did not contain a field for gender of the victim.

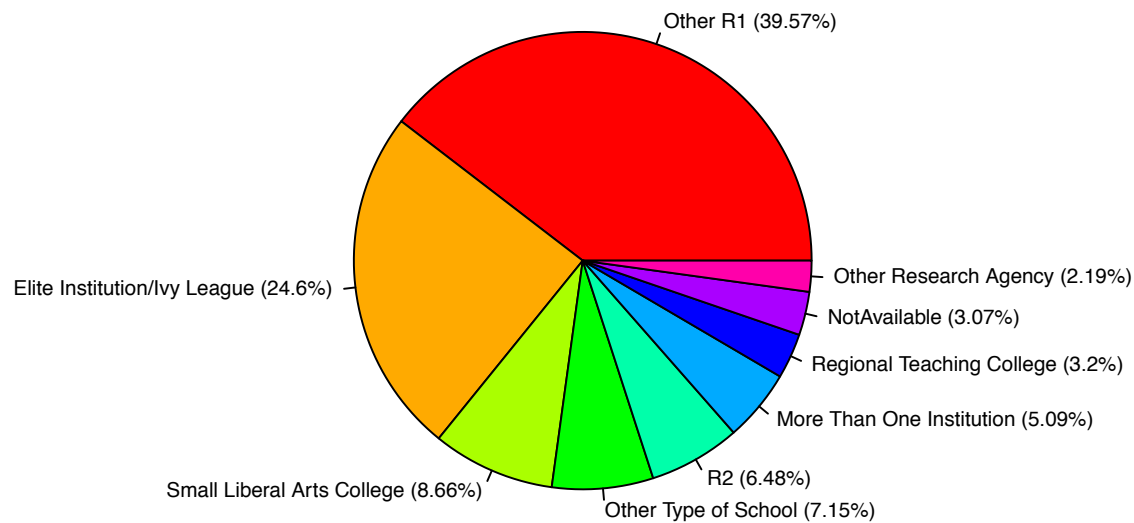

Figure 8: Distribution of Types of Institutions

Considering the fields of study[4,5], we categorized the disciplines into six categories including social sciences and humanities, natural sciences, applied and formal, classics, interdisciplinary, and university administration. While 83.56% of the fields were recognizable, the rest of them were empty or were not detectable, such as *"many years later I still don't want to say"* (Figure 10). About two-thirds of the stories were reported by people in the social sciences and humanities field, again likely reflecting the population of people who utilized the survey. Due to the very low number of records in the classics, interdisciplinary, and university administration categories, we compared the social sciences and humanities, natural sciences, and applied & formal categories based on the average weight of the detected topics. Our results show that there is no significant difference (P-value >0.05) between the topics considering these three categories.

---

[4] [https://en.wikipedia.org/wiki/Branches_of_science](https://en.wikipedia.org/wiki/Branches_of_science)
[5] [https://en.wikipedia.org/wiki/Portal:Science/Categories_and_Main_topics](https://en.wikipedia.org/wiki/Portal:Science/Categories_and_Main_topics)

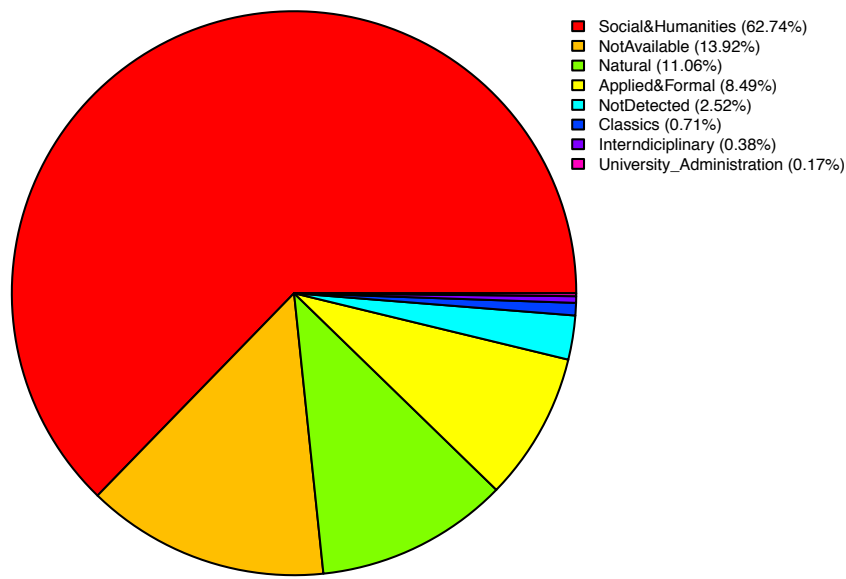

Figure 10: Distribution of Fields of Study

## 5. Discussion

The Internet and social media are bringing many previously "hidden" topics, such as sexual harassment, into the open. The large volume of this data necessitates computational methods, such as text mining, to find the patterns in the data. This study used a mixed methods approach, first using text mining to identify patterns in the data, and then human coders utilizing qualitative coding to make sense of the text mining results. For the qualitative analyses, we used the categories of sex discrimination and sexual harassment previously identified in the literature to guide our coding and grouping of the topics. Furthermore, an outside coder demonstrated very good reliability with our coding. Thus, we feel confident that the sexual harassment themes found in this study are representative of the stories posted in the sexual harassment survey.

The themes we found in the data are consistent with the general sexual harassment literature. There are some aspects of the academic setting, however, that differentiate this setting from many non-academic workplaces. The first is the large power differential between faculty and students. Students depend on faculty in their field in a way that few employees depend on their bosses, for teaching them the information they need to advance in their field, giving them grades that determine what kind of future career they will have, writing letters of recommendation, providing career development opportunities such as research experience, mentoring, networking, funding, professional opportunities, and introducing them to important people in their field. This high level of dependence on an academic advisor is particularly pronounced for graduate students. Tenured professors, in turn, have a level of job security that is unmatched in virtually any other profession. For professors who choose to abuse their power, the power differential between them and those below them in the academic hierarchy can lead to a highly sexualized environment, characterized by unwanted sexual attention towards students in the research lab, classroom, professor's office, academic conferences, field sites, or study-abroad programs.

Unwanted sexual attention had the largest share of topics (23 out of the 41 topics), and seven of the top ten highest weighted topics, indicating that this form of harassment was particularly prevalent among the academic environments of the survey respondents.

The invulnerability of tenured professors can lead to what has been referred to as the sexual supermarket (Fitzgerald & Weitzman, 1990). In this environment, some professors behave as if one of the privileges of their job is to hit on students. In the subtheme *Professor-student affairs*, professors sought out students to date or have affairs with. Professors who sexually harassed students typically faced very few, if any, consequences. As survey respondents noted in the stories in the *Retaliation* theme, the person who paid the heaviest price for the harassment was, almost always, the victim of the harassment.

The survey data contained information about gender of harasser, type of institution at which the harassment occurred, and the field of study of the victim. The vast majority (90%) of harassers were male. We found no differences in harassment topics when we compared the gender of harassers (male or female) and different academic disciplines (social & humanities, natural sciences, and applied & formal).

While most of the sexual harassment experiences occurred at R1 and elite institution/Ivy league universities, this likely reflects the population of academics who utilized theprofessorisin.com website. Differences in certain topics of sexual harassment by type of institution were found. Regional teaching colleges had a greater proportion of T1 (attempts to turn professional interactions into romantic or sexual interactions) than R1 universities. R2 universities had a greater proportion of T32 (predatory professors harassing students in the classroom) than the R1 universities and the elite institution/Ivy schools. Regional teaching colleges had a greater proportion of T1 (attempts to turn professional interactions into romantic or sexual interactions) and T20 (sexual comments and unwanted touching) than elite institution/Ivy league schools and small liberal arts schools. Finally, T7 (professor getting student alone so he can harass her), and T20 (Sexual comments, unwanted touching) occurred more frequently at regional teaching colleges than R2 universities.

In sum, moving from institution type category 1 (elite/Ivy league school), 2 (R1), 3 (R3), and 4 (small liberal arts colleges) to category 5 (regional teaching colleges) represented a spectrum from high research activity to less research activity universities. The four topics that significantly varied by institution type, T1, T7, T20, and T32, were all under the unwanted sexual attention theme. These findings revealed that these four unwanted sexual attention topics had an increasing trend from high research activity to less research activity universities. However, none of the other topics differed by type of institution.

## 6.Conclusion

Sexual harassment in academia is often a hidden problem, because victims do not usually report their experiences. Recent social movements have encouraged people to post their personal experiences on the web anonymously, providing an opportunity for researchers to gain new knowledge by analyzing these experiences. However, analyzing a large number of online comments in text format is a time-consuming and labor-intensive process that is greatly aided by computational methods. Applying text mining, this research investigates sexual harassment experiences in academia posted on a web platform survey.

This paper utilizes an efficient approach to provide a better understanding of sexual harassment in academia. This study detects and categorizes topics, and explores their variation by aggregating the weight of topics across the harasser gender, institution type, and victim's field of study. This study recognizes 41 sexual harassment related topics, ranks them based on their weight, and categorizes them in five themes. While there is not a significant difference between the weight of topics aggregated on the gender of harasser and victim's field of study, type of institute did show differences, with several unwanted sexual attention topics tending to be weighted more highly at regional teaching colleges than other types of institutions. Our findings demonstrate that text mining is a useful method to investigate numerous sexual harassment experiences.

While this research provides insight into the problem of sexual harassment in academia, it has some limitations. First, we were unable to find some information, such as gender of the victims. We also do not know age, location, or other demographic information for the victims or harassers. Second, we collected data from a single web resource. To address these limitations, future research may consider different data sources, detect and investigate more information about victims and harassers, or use computational methods to infer demographic information from the data sources.

This research may assist relevant researchers for further investigation of this paper's dataset, and for exploring other sexual harassment issues utilizing other large datasets. The information may also be useful to academic institutions for (1) improving existing sexual harassment policies by developing targeted prevention and support policies and programs, (2) initiating discussions about sexual harassment among students, faculties, and employees, perhaps using some of the stories from this dataset as examples, (3) promoting gender equality, and (4) encouraging lawmakers to propose new regulations to create a safe academic environment.

## 6. Conflict of interest

The authors state that they have no conflict of interest.

### Acknowledgment

This work is partially supported by an Advanced Support for Innovative Research Excellence (ASPIRE) grant (15700-18-47805) from the Office of the Vice President for Research at the University of South Carolina. All opinions, findings, conclusions and recommendations in this paper are those of the authors and do not necessarily reflect the views of the funding agency.